\documentclass[aps,prb,twocolumn,superscriptaddress,showpacs,amsmath,amssymb,amsfonts,floatfix]{revtex4}

\usepackage{graphicx}
\usepackage{bm}

\begin{document}

\title{Chiral magnetic effect in two-band lattice model of Weyl semimetal}

\author{Ming-Che Chang}
\affiliation{Department of Physics, National Taiwan Normal
University, Taipei, Taiwan}

\author{Min-Fong Yang}
\email{mfyang@thu.edu.tw}
\affiliation{Department of Applied Physics,
Tunghai University, Taichung, Taiwan}

\date{\today}

\begin{abstract}
Employing a two-band model of Weyl semimetal, the existence of the chiral magnetic effect (CME) is established within the linear-response theory. The crucial role played by the limiting procedure in deriving correct transport properties is clarified. Besides, in contrast to the prediction based on linearized effective models, the value of the CME coefficient in the uniform limit shows nontrivial dependence on various model parameters. Even when these parameters are away from the region of the linearized models, such that the concept of chirality may not be appropriate, this effect still exists. This implies that the Berry curvature, rather than the chiral anomaly, provides a better understanding of this effect.
\end{abstract}


\maketitle

\section{Introduction}

Weyl semimetals are gapless systems with nontrivial momentum-space topology.~\cite{Volovik} Their low-energy excitations are described by three-dimensional (3D) Weyl fermions emerging from the degenerate points between energy bands, the so-called Weyl nodes, in momentum space. In lattice systems, Weyl nodes with opposite chiralities always appear in pairs because of the Nielsen-Ninomiya theorem.~\cite{Nielsen_Ninomiya} These nodes are topologically protected in the sense that they are robust against small perturbations unless two Weyl nodes with opposite chiralities annihilate each other. Interestingly, Weyl nodes act as (anti-)monopoles of the Berry curvature with unit strength. They thus are sources of many remarkable electromagnetic responses, such as chiral anomaly and other anomaly-induced transport phenomena.~\cite{Hosur_Qi2013} Since Weyl semimetals  could have these novel effects, material compounds with such nodal points are soon proposed, and some of them verified in experiments.~\cite{Wan_etal2011,Burkov_Balents2011,Xu_etal2011}
A closely related class of materials with two Weyl nodes at the same location has also been under intense study recently.~\cite{Borisenko_etal2014,Liu_etal2014,Neupane_etal2014,Jeon_etal2014} More of these nodal materials, with possible applications, are likely to emerge in the near future.

Besides the anomalous Hall effect,~\cite{Burkov_Balents2011,
Zyuzin_Burkov2012,Grushin2012,Goswami_Tewari2013} Weyl semimetals are proposed to have the chiral magnetic effect (CME), if pairs of Weyl nodes have different energies.~\cite{Zyuzin_Burkov2012,
Goswami_Tewari2013,Stephanov_Yin2012,Son_Yamamoto2012,Zyuzin_etal2012} This effect gives a dissipationless electric current $\mathbf{J}$ flowing along the direction of an applied magnetic field $\mathbf{B}$.~\cite{Fukushima_etal2008} In the simplest case of a Weyl semimetal with just a single pair of Weyl nodes, employing the low-energy effective theory with unbounded linear dispersion, the chiral magnetic current is shown to be $\mathbf{J}=-\alpha\mathbf{B}$ with the CME coefficient $\alpha=2 b_{0}e^{2}/h^2$, where $2b_0$ is the energy separation between Weyl nodes. Notably, $\alpha$ takes a universal form and is independent of chemical potential or temperature. This result is usually explained by its resemblance to the chiral anomaly.~\cite{Hosur_Qi2013}

However, unlike the prediction of the anomalous Hall effect in Weyl semimetals, there has been an active debate regarding the existence of the CME. It was shown that, for the unbounded linearized effective models, the result could depend crucially on the regularization scheme.~\cite{Basar_etal2014,Landsteiner2014} (See also the discussion presented in Appendix~\ref{cutoff}.) These authors emphasize that it is the ultra-violet cutoff in \emph{energy}, rather than in \emph{momentum}, that should be used in condensed matter. Moreover, if energy cutoff is employed, then the CME will be absent.

In a recent work based on a lattice model, where the artificial ultra-violet cutoff is not necessary, the equilibrium chiral magnetic current is shown to be zero also.~\cite{Vazifeh_Franz2013} Indeed, as explained by Ba\c{s}ar \emph{et al.}, a nonzero equilibrium chiral magnetic current induced by a static magnetic field at zero temperature would cause a conceptual problem.~\cite{Basar_etal2014}
Once it exists, it can be used to extract energy from a ground state in global equilibrium at zero temperature, which should be impossible.
Therefore, the existence of the CME seems to be ruled out.

Later, the absence of the CME was challenged in Ref.~\onlinecite{Chen_etal2013}, in which the importance of the limiting procedure around zero wave-vector ($\mathbf{q}=0$) and zero frequency ($\omega=0$) is pointed out (see also Ref.~\onlinecite{Goswami2013}). Notice that the previous null result was achieved under static magnetic fields,~\cite{note1} which corresponds to the \emph{static} limit (i.e., $\omega=0$ before $\mathbf{q}\to 0$).
Physically, the absence of current in this limit is expected, because a new equilibrium distribution of charges will be established and after that no current will flow.~\cite{Mahan,Thouless,Luttinger1964,note2} Besides, the result of no chiral magnetic current in the static limit is consistent with the aforementioned argument in Ref.~\onlinecite{Basar_etal2014}, because this limit gives equilibrium properties. Therefore, the correct dc transport properties should be calculated under the \emph{uniform} (or long-wavelength) limit (i.e., $\mathbf{q}=0$ before $\omega\to 0$). Further justification of the uniform limit in deriving the dc transport quantities is elucidated in Refs.~\onlinecite{Shastry2009,Peterson_Shastry2010,Tremblay,note3}.

In this paper, we try to provide a definite answer to the existence of the CME. The controversies in literature are clarified as best as we can. For simplicity, we focus on a generic two-band model of Weyl semimetals. Because the main feature of Weyl semimetals is the existence of point degeneracies (Weyl points) between energy bands, a simple two-band model is sufficient for the essential physics of Weyl semimetals. Within the linear-response theory, we arrive at distinctive expressions of the CME coefficient for two different limits approaching zero frequency-momentum. It shows that the limiting procedure does play a crucial role in deriving correct transport properties, in favour of the arguments in Ref.~\onlinecite{Chen_etal2013}. Since these results are based on a lattice model, where no linearization in energy dispersion is employed and no artificial ultra-violet cutoff is introduced, our conclusion should have general validity.

To evaluate the magnitude of the CME coefficient numerically, a specific form of a minimal lattice model with only two Weyl nodes is chosen. We find that, in the static limit, there is no CME; while the CME coefficient does become nonzero in the uniform limit. The former is consistent with the general argument discussed in the early literature.~\cite{Mahan,Thouless,Luttinger1964} Besides, in contrast to the prediction based on linearized effective models, the value of the CME coefficient in the uniform limit shows nontrivial dependence on various model parameters, including the chemical potential, the temperature, and the energy separation between two Weyl nodes. In particular, when the chemical potential lies far away from the energies of the two Weyl nodes, such that linearized models are no longer appropriate, the CME coefficient can still be finite [see Fig.~\ref{fig:t1_dep}(b) for large $t_1$]. This nonzero value is hard to be understood from the perspective of chiral anomaly, since the concept of chirality becomes ambiguous there. This implies that it is the Berry curvature, rather than the chiral anomaly, that might provide a better conceptual framework of this effect.

This paper is organized as follows: In Sec.~II, both the static and the uniform limits of the CME coefficients for a generic two-band model are derived within the linear-response theory. In Sec.~III, the behavior of the CME coefficient and the anomalous Hall conductivity  under the variation of different model parameters is investigated numerically. We summarize our work in Sec.~IV. In App.~A, the subtlety of using ultra-violet cutoff for a linearized effective model is explained.
Finally, using the semiclassical wave-packet dynamics, the correct forms of the distribution function in both of the zero frequency-momentum limits are discussed in App.~B.

\section{linear-response theory for generic two-band model}
\label{Sec:LRT}

For simplicity, we consider a generic two-band model of Weyl semimetal, which is sufficient to describe its main feature of point degeneracies between energy bands. Within the linear-response theory, both expressions of the CME coefficient $\alpha(\mathbf{q},\omega)$ in the uniform and the static limits are derived, as shown in Eqs.~\eqref{chmag1} and \eqref{chmag2} [or \eqref{chmag3}]. They are the major results in this section. It shows that the limiting procedure does play a crucial role in deriving correct transport properties, supporting the discussions in Ref.~\onlinecite{Chen_etal2013}. That is, $\alpha(\mathbf{q},\omega)$ behaves non-analytically in the zero frequency-momentum limit. Readers not interested in this derivation may skip directly to Section~III, wherein the behavior of the CME coefficient under the variation of several model parameters is discussed.

The most general two-band Hamiltonian describing a
non-interacting system can be expressed in the following form:
\begin{equation}\label{Hami}
H= \sum_\mathbf{k} H(\mathbf{k}),\quad
H(\mathbf{k})=\epsilon(\mathbf{k})
+ \mathbf{d}(\mathbf{k})\cdot\bm{\sigma} \; ,
\end{equation}
where $\sigma^\alpha$ ($\alpha=x$, $y$, $z$) are the three Pauli matrices
and $\mathbf{k}$ stands for the Bloch wavevector of electrons. The two bands may represent different physical degrees of freedom (say, spin or pseudospin for two sublattices) depending on the context. The band energies for the present model are
\begin{equation}
E_\pm(\mathbf{k})=\epsilon(\mathbf{k})\pm d(\mathbf{k}) \; ,
\end{equation}
where $d(\mathbf{k})=\sqrt{\mathbf{d}(\mathbf{k})\cdot\mathbf{d}(\mathbf{k})}$ is the norm of the 3-vector $\mathbf{d}(\mathbf{k})$.

Within the linear-response theory, the expectation value of electric current becomes
\begin{equation}
J^i(\mathbf{q},\omega) = \Pi_{ij}(\mathbf{q},\omega)
A^j(\mathbf{q},\omega) \; ,
\end{equation}
where the vector potential $\mathbf{A}(\mathbf{q},\omega)$ is related to the applied magnetic field $\mathbf{B}(\mathbf{q},\omega)$ by
$\mathbf{B}(\mathbf{q},\omega)=i\mathbf{q}\times\mathbf{A}(\mathbf{q},\omega)$.
Here the Einstein summation convention for the repeated indices is adopted. $\Pi_{ij}(\mathbf{q},\omega)$ is the \emph{retarded}
current-current correlation function, which can be calculated by the formula,
\begin{eqnarray}
\Pi_{ij}(\mathbf{q},i\nu_m) %
&=&\frac{1}{V\beta} \sum_{\mathbf{k},n} \mathrm{tr}
\left[\hat{J}_i(\mathbf{k})
G(\mathbf{k}+\mathbf{q},i\omega_n+i\nu_m)\right. \nonumber\\
&&\qquad\left.\times \hat{J}_j(\mathbf{k})G(\mathbf{k},i\omega_n)\right].
\label{kubo}
\end{eqnarray}
Here $V$ is the volume of the system, $\beta=1/k_B T$ is the inverse temperature, and $\omega_n=(2n+1)\pi/\beta$ and $\nu_m=2m\pi/\beta$ are the fermionic and the bosonic Matsubara frequencies, respectively. For the two-band model in Eq.~\eqref{Hami}, the electric current operator becomes
\begin{equation}
\hat{J}_i(\mathbf{k}) %
=-\frac{e}{\hbar}\frac{\partial H(\mathbf{k})}{\partial k_i} %
=-\frac{e}{\hbar} \left[
\frac{\partial \epsilon(\mathbf{k})}{\partial k_i} +
\frac{\partial \mathbf{d}(\mathbf{k})}{\partial k_i}\cdot\bm{\sigma}
\right]   \label{currentoperator}
\end{equation}
and the Matsubara Green function takes the form
\begin{equation}
G(\mathbf{k},i\omega_n)
=\frac{P_+(\mathbf{k})}{i\omega_n+\mu-E_+(\mathbf{k})}
+\frac{P_-(\mathbf{k})}{i\omega_n+\mu-E_-(\mathbf{k})} \; .
\label{Greenfunction}
\end{equation}
Here $P_\pm(\mathbf{k}) = [ 1 \pm \hat{\mathbf{d}}(\mathbf{k})\cdot\bm{\sigma}]/2$ is the projection operator with $\hat{\mathbf{d}}(\mathbf{k})=\mathbf{d}(\mathbf{k})/d(\mathbf{k})$ being the unit vector along the direction of $\mathbf{d}(\mathbf{k})$. Note that $\hat{\mathbf{d}}(\mathbf{k})$ is singular if the norm $d(\mathbf{k})$ vanishes at a ${\bf k}$ point.

Because the electric current is a vector and the magnetic field is a pseudovector, the CME coefficient is parity-odd. We thus focus only on the contribution of the \emph{anti-symmetric} part (or the parity-odd part) of $\Pi_{ij}$, which is defined by
\begin{equation}\label{Pi_anti_def}
\Pi_{ij}^\mathrm{anti}(\mathbf{q},\omega) \equiv i \alpha(\mathbf{q},\omega) \epsilon^{ijk} q_k \; ,
\end{equation}
such that
\begin{eqnarray}
J^i_\mathrm{CME}(\mathbf{q},\omega)
&=& i \alpha(\mathbf{q},\omega) \epsilon^{ijk} q_k A^j(\mathbf{q},\omega) \nonumber \\
&=& -\alpha(\mathbf{q},\omega) B^i(\mathbf{q},\omega) \; .
\end{eqnarray}
Therefore, the zero frequency-momentum limit of $\alpha(\mathbf{q},\omega)$ gives the CME coefficient $\alpha$ for a uniform, static magnetic field.

We consider the case in which the current flows along $x$ direction and $\mathbf{q}=q\hat{\mathbf{z}}$. Thus $\alpha(\mathbf{q},\omega)$ can be obtained by
\begin{equation}\label{alpha}
\alpha(\mathbf{q},\omega) = -\frac{i}{q}\Pi_{xy}^\mathrm{anti}(\mathbf{q},\omega)  \; .
\end{equation}
To calculate $\Pi_{xy}^\mathrm{anti}(\mathbf{q},\omega)$, we first substitute Eqs.~\eqref{currentoperator} and \eqref{Greenfunction} into Eq.~\eqref{kubo}. After performing the Matsubara sum over $i\omega_n$
and then making analytic continuation to the real frequency, $i\nu_m \to \hbar\omega+i\delta$, one has
\begin{eqnarray}
&&\Pi_{xy}(\mathbf{q},\omega) \nonumber\\%
&&=\frac{1}{V}\sum_{s,t=\pm}\sum_{\mathbf{k}}%
\mathrm{tr}\left[\hat{J}_{x}(\mathbf{k})P_s(\mathbf{k}+\mathbf{q})\hat{J}_y(\mathbf{k})P_t(\mathbf{k})\right]
\times \nonumber\\
&&\qquad\qquad\qquad
\frac{f_t(\mathbf{k})-f_s(\mathbf{k}+\mathbf{q})}%
{\hbar\omega+i\delta+E_t(\mathbf{k})-E_s(\mathbf{k}+\mathbf{q})} \; ,
\end{eqnarray}
where $f_t(\mathbf{k})=1/\{e^{\beta[E_t(\mathbf{k})-\mu]}+1\}$ is the Fermi-Dirac distribution function for band $t$.

Therefore, the antisymmetric part of $\Pi^\textrm{anti}_{xy}(\mathbf{q},\omega)\equiv
\frac{1}{2} [\Pi_{xy}(\mathbf{q},\omega)-\Pi_{yx}(\mathbf{q},\omega)]$ becomes
\begin{eqnarray}\label{Pi_anti}
&&\Pi^\textrm{anti}_{xy}(\mathbf{q},\omega)
= \frac{1}{V}\sum_{s,t=\pm}\sum_{\mathbf{k}}%
M_{t,s}(\mathbf{k}; \mathbf{q}) \times \nonumber\\%
&&\qquad\qquad\qquad %
\frac{f_t(\mathbf{k})-f_s(\mathbf{k}+\mathbf{q})}%
{\hbar\omega+i\delta+E_t(\mathbf{k})-E_s(\mathbf{k}+\mathbf{q})}  \; ,
\end{eqnarray}
where
\begin{eqnarray}\label{M}
&&M_{t,s}(\mathbf{k}; \mathbf{q}) \nonumber\\
&&=\frac{1}{2} \mathrm{tr}\left[\hat{J}_{x}(\mathbf{k})P_s(\mathbf{k}+\mathbf{q})\hat{J}_y(\mathbf{k})P_t(\mathbf{k})\right] - (x\leftrightarrow y) \nonumber\\
&&=\frac{1}{2}\frac{e^2}{\hbar^2}\;i\epsilon^{\alpha\beta\gamma}%
\left\{\frac{\partial d_\alpha(\mathbf{k})}{\partial k_x}
\frac{\partial d_\beta(\mathbf{k})}{\partial k_y} %
\left[t\hat{d}_\gamma(\mathbf{k})-s\hat{d}_\gamma(\mathbf{k}+\mathbf{q})\right]
\right. \nonumber \\
&&\quad + \left.
\left[\frac{\partial \epsilon(\mathbf{k})}{\partial k_y} %
\frac{\partial d_\alpha(\mathbf{k})}{\partial
k_x}-\frac{\partial \epsilon(\mathbf{k})}{\partial k_x} %
\frac{\partial d_\alpha(\mathbf{k})}{\partial k_y}\right] \times \right. \nonumber \\
&&\qquad %
\left. s\hat{d}_\beta(\mathbf{k}+\mathbf{q}) t\hat{d}_\gamma(\mathbf{k})
\right\} \; .
\end{eqnarray}
Notice that only the three-$\sigma^\alpha$ terms in the expansion give non-vanishing contributions to the trace.

Now the origin of the non-analyticity of $\alpha(\mathbf{q},\omega)$ in the zero frequency-momentum limit can be understood as follows. The sum over band indices in Eq.~\eqref{Pi_anti} can be separated into two parts: the intraband contribution for $s=t$ and the interband contribution for $s=-t$. We find that, when $s=t$, $M_{t,s=t}(\mathbf{k}; \mathbf{q})$ vanishes at $\mathbf{q}=q\hat{\mathbf{z}}=0$. Thus the leading term in small $q$ becomes
\begin{equation}
M_{t,s=t}(\mathbf{k}; \mathbf{q}) %
\simeq q \left. \frac{\partial M_{t,s=t}}{\partial q}\right|_{q=0} \; .
\end{equation}
On the other hand, for $s=t$ and $q\ll 1$,
\begin{eqnarray}
&&\frac{f_t(\mathbf{k})-f_t(\mathbf{k}+\mathbf{q})}%
{\hbar\omega+i\delta+E_t(\mathbf{k})-E_t(\mathbf{k}+\mathbf{q})} %
\nonumber \\
&&\approx -\frac{\partial f_t}{\partial E_t}\; \frac{\mathbf{q}\cdot\nabla_\mathbf{k}E_t(\mathbf{k})}%
{\hbar\omega+i\delta-\mathbf{q}\cdot\nabla_\mathbf{k}E_t(\mathbf{k})}
\; .
\end{eqnarray}
Employing these observations and Eqs.~\eqref{alpha} and \eqref{Pi_anti}, we find that the \emph{intraband} contribution of $\alpha(\mathbf{q},\omega)$ vanishes identically in the uniform limit ($\mathbf{q}=0$ before $\omega\to 0$), which corresponds to the dc limit of a transport coefficient. However, it becomes nonzero in the static limit ($\omega=0$ before $\mathbf{q}\to 0$). As a consequence, the two limits, $\lim_{\mathbf{q}\to 0}$ and $\lim_{\omega\to 0}$, do not commute and  $\alpha(\mathbf{q},\omega)$ thus behaves non-analytically at $(\mathbf{q}, \omega)=(\mathbf{0}, 0)$. This conclusion remains true even after including the interband contribution of $\alpha(\mathbf{q},\omega)$, since the latter has a unique value in both frequency-momentum limits.

After lengthy calculations, the intraband contribution of $\alpha(\mathbf{q},\omega)$ in small $q$ is found to be
\begin{eqnarray}\label{intraband}
&&\alpha(\mathbf{q},\omega)|_\textrm{intra} \nonumber\\%
&&\approx\frac{e^2}{\hbar^2}\frac{1}{V}\sum_{t=\pm}\sum_{\mathbf{k}}
\frac{\partial f_t}{\partial E_t}\; %
\frac{q \frac{\partial E_t}{\partial k_z}} %
{\hbar\omega+i\delta-q \frac{\partial E_t}{\partial k_z}} \times \nonumber\\
&&\quad t\left\{
\frac{\partial E_t(\mathbf{k})}{\partial k_x}\Omega^x_{t}(\mathbf{k})
+\frac{\partial E_t(\mathbf{k})}{\partial k_y}\Omega^y_{t}(\mathbf{k}) \right\} d(\mathbf{k})  \; ,
\end{eqnarray}
and the interband part in the zero frequency-momentum limit is
\begin{eqnarray}
&&\alpha(\mathbf{q}=\mathbf{0},\omega=0)|_\textrm{inter} \nonumber\\
&&=\frac{e^2}{\hbar^2}\frac{1}{V}\sum_{t=\pm}\sum_{\mathbf{k}}
\left[ f_t(\mathbf{k})
\frac{\partial \epsilon(\mathbf{k})}{\partial \mathbf{k}}\cdot\mathbf{\Omega}_{t}(\mathbf{k}) \right. \nonumber\\
&&\quad \left. %
-t\frac{\partial f_t}{\partial E_t}%
\frac{\partial E_t(\mathbf{k})}{\partial k_z}
\Omega^z_{t}(\mathbf{k}) d(\mathbf{k}) \right]  \; .
\end{eqnarray}
Here the Berry curvatures $\Omega^i_{\pm}(\mathbf{k})$ ($i=x$, $y$, $z$) are given by the formula~\cite{Bernevig}
\begin{eqnarray}\label{Berry}
\Omega^i_{\pm}(\mathbf{k})= %
\pm \epsilon^{ij\ell} \frac{1}{4d^3(\mathbf{k})} \; \mathbf{d}(\mathbf{k}) \cdot \left[ \frac{\partial \mathbf{d}(\mathbf{k})}{\partial k_j}\times \frac{\partial \mathbf{d}(\mathbf{k})}{\partial k_\ell}\right] \; .
\end{eqnarray}

Finally, after combining the intraband part with the interband part,
the CME coefficient in the uniform limit becomes
\begin{eqnarray}\label{chmag1}
&&\lim_{\omega\to 0} \lim_{q\to 0} \alpha(\mathbf{q},\omega) \nonumber\\
&&=\frac{e^2}{\hbar}\int \frac{d^3k}{(2\pi)^3} \; \sum_{t=\pm} \left[\frac{\mathbf{v}_{\mathbf{k},+}+\mathbf{v}_{\mathbf{k},-}}{2} \cdot \mathbf{\Omega}_{\mathbf{k},t} \; f_t(\mathbf{k}) \right. \nonumber\\%
&&\qquad\qquad\qquad \left. %
-\frac{1}{3} t\;d(\mathbf{k}) \; \mathbf{v}_{\mathbf{k},t} \cdot \mathbf{\Omega}_{\mathbf{k},t} \; \frac{\partial f_t}{\partial E_t}\right] \; ,
\end{eqnarray}
where $\mathbf{v}_{\mathbf{k},\pm}=(1/\hbar)\nabla_\mathbf{k}E_\pm(\mathbf{k})$ are the group velocities and we have symmetrized over three spatial directions. Notice that, as seen from Eq.~\eqref{intraband}, the intraband part vanishes in this limit and only the \emph{interband} part contributes to the CME coefficient. However, both the intraband and the interband parts are nonzero in the static limit, and a different formula is thus reached,
\begin{eqnarray}\label{chmag2}
&&\lim_{q\to 0} \lim_{\omega\to 0} \alpha(\mathbf{q},\omega) \nonumber\\
&&=\frac{e^2}{\hbar}\int \frac{d^3k}{(2\pi)^3} \; \sum_{t=\pm} \left[\frac{\mathbf{v}_{\mathbf{k},+}+\mathbf{v}_{\mathbf{k},-}}{2} \cdot \mathbf{\Omega}_{\mathbf{k},t} \; f_t(\mathbf{k}) \right. \nonumber\\%
&&\qquad\qquad\qquad \left. %
- t\;d(\mathbf{k}) \; \mathbf{v}_{\mathbf{k},t} \cdot \mathbf{\Omega}_{\mathbf{k},t} \; \frac{\partial f_t}{\partial E_t}\right] \; .
\end{eqnarray}
After integration by parts for the second term, Eq.~\eqref{chmag2} can be rewritten as
\begin{equation}\label{chmag3}
\lim_{q\to 0} \lim_{\omega\to 0} \alpha(\mathbf{q},\omega)
=\frac{e^2}{\hbar}\int \frac{d^3k}{(2\pi)^3} \; \sum_{t=\pm} \mathbf{v}_{\mathbf{k},t}\cdot\mathbf{\Omega}_{\mathbf{k},t} \; f_t(\mathbf{k})  \; ,
\end{equation}
which is nothing but the result obtained in the semiclassical approach for a static magnetic field.~\cite{Zhou_etal2013,Basar_etal2014,Landsteiner2014}

In the next section, an explicit form of the two-band model is considered. We find that either Eq.~\eqref{chmag2} or Eq.~\eqref{chmag3} always gives a vanishing value of the CME coefficient for all model parameters. It shows that there is no CME in the static limit, which is consistent with the general argument mentioned in Refs.~\onlinecite{Mahan,Thouless,Luttinger1964}.

Before closing this section, some remarks are in order.
First, the same approach has been employed in Ref.~\onlinecite{Goswami2013}. While our result in the uniform limit [i.e., Eq.~\eqref{chmag1}] agrees with theirs, our expression in the static limit [i.e., Eq.~\eqref{chmag2}] does not. We believe that the expression here is correct, because it agrees with the one in the semiclassical approach for a static magnetic field [i.e., Eq.~\eqref{chmag3}].

Second, one should not be too surprised by the nonanalyticity of
$\alpha(\mathbf{q},\omega)$ at $(\mathbf{q}, \omega)=(\mathbf{0}, 0)$, because many response functions are known to behave non-analytically in the neighborhood of zero momentum and zero frequency.
For instance, the static current-current correlation function $\chi_{xx}(\mathbf{q},\omega=0)$ of superfluids will give the normal fluid density $\rho_n$ under the limiting procedure $q_x\to 0$ being taken before $q_y$, $q_z\to 0$.~\cite{Baym,Forster,Huang} However, it  yields the total density $\rho$ if $q_y$, $q_z\to 0$ are taken before $q_x\to 0$.
Besides, the dielectric response function $\epsilon(\mathbf{q},\omega)$ of an electron gas behaves very differently in the static and the uniform limits.~\cite{Kittel} The former, $\epsilon(\mathbf{q},\omega=0)$, describes the electrostatic screening of electric fields, while the latter, $\epsilon(\mathbf{q}=0,\omega)$, gives the plasma oscillation in the uniform limit.
The non-analyticity in the correlation functions appears as well in the field theories with Lorentz symmetry at finite temperatures. As an example, the induced Chern-Simons coefficient of (2+1)-dimensional relativistic field theory (QED), which is extracted from the parity violating part of current-current correlation function of fermions, has been shown to behave non-analytically in the zero frequency-momentum limit at finite temperature.~\cite{Kao_Yang1993}
All these facts show that the order of the limits is often important and only physics context can dictate the proper order.

\section{dependence of CME on model parameters}

In order to have a concrete understanding of the formal results in Sec.~\ref{Sec:LRT}, we adopt a minimum model with only two Weyl nodes to investigate the CME. It is an extension of the Qi-Wu-Zhang model in the study of the two-dimensional (2D) quantum anomalous Hall effect~\cite{Qi2006} and is equivalent to the 3D lattice model considered in Ref.~\onlinecite{Ramamurthy_Hughes14}. The Hamiltonian is
\begin{equation}\label{H}
H(\mathbf{k})=t_1\cos k_z+H_{so}+H_m \; ,
\end{equation}
where the spin-orbit interaction and the magnetization-related parts are
\begin{eqnarray}
H_{so}&=&t_{so}\left( \sin k_x \sigma^x+\sin k_y \sigma^y+\sin k_z \sigma^z \right) \; , \cr
H_m&=&\left( m+2-\cos k_x-\cos k_y \right) \sigma^z \; .
\end{eqnarray}
Here $t_1$ represents the hopping integral along the $z$ direction, $t_{so}$ is the strength of the spin-orbit coupling, and $m$ is the magnetization. Notice that a nonzero $t_1$ would break the space-inversion symmetry about $\mathbf{k}=(\pi/2)\hat{\mathbf{z}}$: $H(\mathbf{k})\to\sigma^{z}H(-\mathbf{k}+\pi\hat{\mathbf{z}})\sigma^{z}$, while $H_m$ breaks time-reversal symmetry: $H(\mathbf{k})\to\sigma^{y}H^*(-\mathbf{k})\sigma^{y}$. The energies of the two bands are
\begin{widetext}
\begin{equation}
E_\pm({\bf k})=t_1\cos k_z
\pm\sqrt{t^2_{so}\left(\sin^2k_x+\sin^2k_y\right)+\left(t_{so}\sin k_z+m+2-\cos k_x-\cos k_y\right)^2} \; .
\end{equation}
\end{widetext}

In addition to the CME, Weyl semimetal exhibits the anomalous Hall effect.~\cite{Burkov_Balents2011,Zyuzin_Burkov2012,
Grushin2012,Goswami_Tewari2013} This effect gives an electric current flowing in a direction perpendicular to the applied electric field. For an electric field in the $x$ direction and a current in the $y$ direction, the  Hall conductivity is
\begin{equation}\label{AHE}
\sigma_H=\frac{e^2}{\hbar}\int \frac{d^3k}{(2\pi)^3} \; \sum_{t=\pm} \Omega^z_t({\bf k}) \; f_t(\mathbf{k})  \; ,
\end{equation}
where $\Omega^z_t({\bf k})$ is the $z$-component of the Berry curvature of band $t$ [see Eq.~\eqref{Berry}]. According to this formula, the anomalous Hall effect can be understood as follows.~\cite{Burkov_Balents2011}  A pair of Weyl nodes with opposite chiralities is connected by a string of gauge singularities (i.e., the Dirac string) in the 3D Brillouin zone (BZ). A 2D cross section of the 3D BZ that intersects the Dirac string would have a quantized 2D Hall conductivity at zero temperature. By summing over the contributions of all the 2D cross sections, a Weyl semimetal would have a nonzero Hall conductivity, and thus shows the anomalous Hall effect.

\begin{figure}
\includegraphics[width=3.2in]{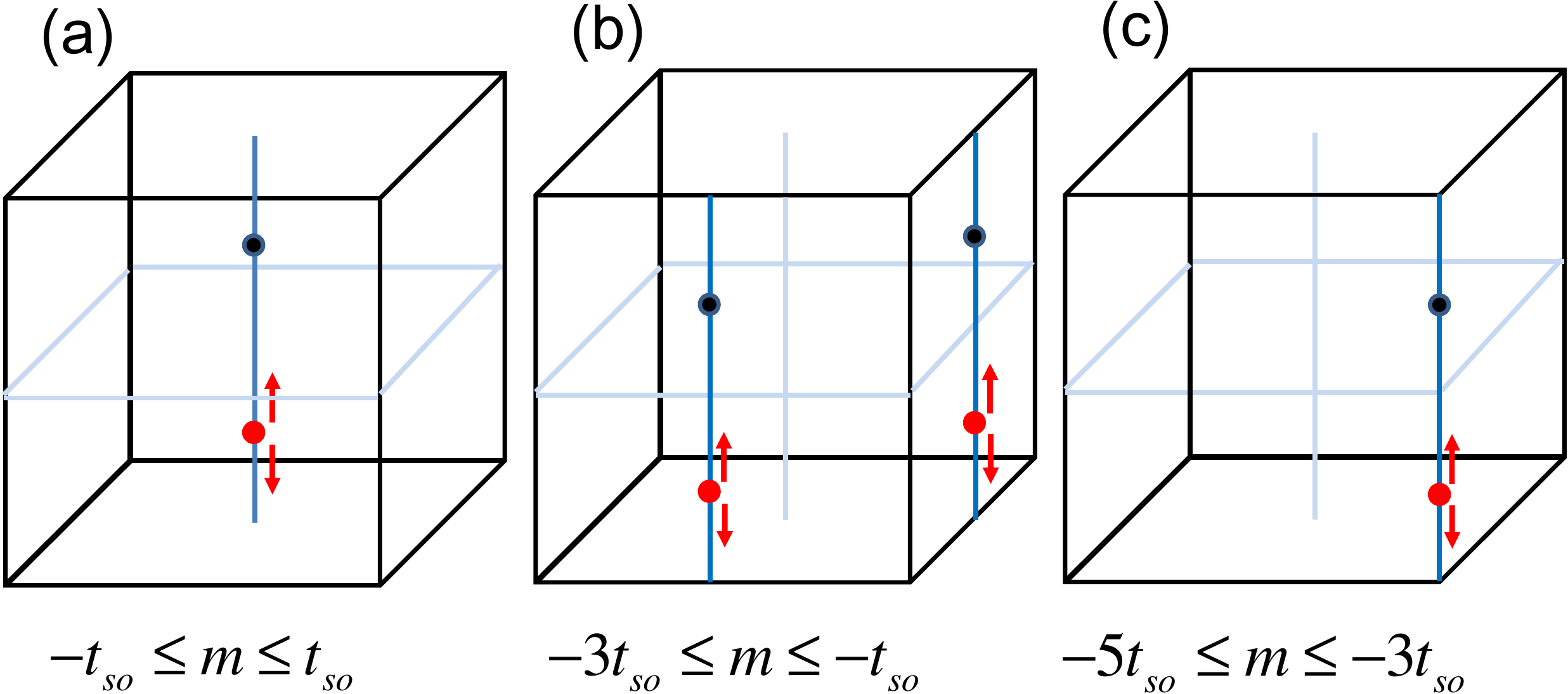}
\caption{Generation and annihilation of Weyl nodes in the 3D Brillouin zone for three different ranges of $m$. (a) Reduce $m$ from $t_{so}$ to $-t_{so}$. Two overlapped Weyl nodes (a Dirac node) first appear at the lower dot of $\mathbf{k}=(0,0,-\pi/2)$, then move away from each other along the directions of the arrows, and finally merge again at the higher dot of $\mathbf{k}=(0,0,\pi/2)$. (b) Reduce $m$ from $-t_{so}$ to $-3t_{so}$. Two Dirac nodes first appear at the two lower dots of $\mathbf{k}=(\pi,0,-\pi/2)$ and $(0,\pi,-\pi/2)$, then each splits to two Weyl nodes along the directions of the arrows, and finally merge again at the two higher dots of $\mathbf{k}=(\pi,0,\pi/2)$ and $(0,\pi,\pi/2)$. (c) Reduce $m$ from $-3t_{so}$ to $-5t_{so}$. Two overlapped Weyl nodes first appear at the lower dot of $\mathbf{k}=(\pi,\pi,-\pi/2)$, then move apart along the directions of the arrows, and merge again at the higher dot of  $\mathbf{k}=(\pi,\pi,\pi/2)$. }
\label{fig:node}
\end{figure}

Depending on the values of $m$ and $t_{so}$ (the latter is assumed to be positive), the ground state can be either in a trivial phase or in an anomalous Hall phase. For $m>t_{so}$, the two bands are separated by an energy gap and the ground state is trivial. When $m=t_{so}$, a Dirac node (i.e., two overlapped Weyl nodes) appears at $\mathbf{k}=(0,0,-\pi/2)$. Upon the decrease in $m$, it splits to two Weyl nodes moving along opposite $k_z$ directions [see Fig.~\ref{fig:node}(a)]. The two nodes merge again at $\mathbf{k}=(0,0,\pi/2)$ when $m=-t_{so}$. Similar to Burkov and Balents' multi-layer model of Weyl semimetal,~\cite{Burkov_Balents2011} the two nodes are linked by a string of gauge singularities. Such a string spans the whole $z$-axis when $m=-t_{so}$ and the system can have the maximum Hall conductivity $e^2/h$ at zero temperature.

In the Burkov-Balents model, when the two nodes annihilate each other, an energy gap is opened, the Hall conductivity remains saturated, and the system enters the quantum anomalous Hall phase. In the present model, however, when one keeps reducing the value of $m$ from $-t_{so}$, two Dirac nodes appear simultaneously at $\mathbf{k}=(\pi,0,-\pi/2)$ and $(0,\pi,-\pi/2)$ [see Fig.~\ref{fig:node}(b)]. Each of the Dirac nodes will split along opposite $k_z$ directions and merge at $\mathbf{k}=(\pi,0,\pi/2)$ and $(0,\pi,\pi/2)$ respectively when $m=-3t_{so}$. Each pair will also stretch out Dirac strings while moving away from each other. However, we see in a numerical calculation later that these two strings are anti-Dirac strings, so that the Hall conductivity is reduced as the strings are stretched.

When $m$ is reduced further from $-3t_{so}$, a Dirac node first appears at $\mathbf{k}=(\pi,\pi,-\pi/2)$, then splits to two Weyl nodes connected by a normal Dirac string, and finally annihilate with each other at $\mathbf{k}=(\pi,\pi,\pi/2)$ [see Fig.~\ref{fig:node}(c)]. After that, the system becomes fully gapped and returns to the trivial phase with zero Hall conductivity.

\begin{figure}
\includegraphics[width=3.3in]{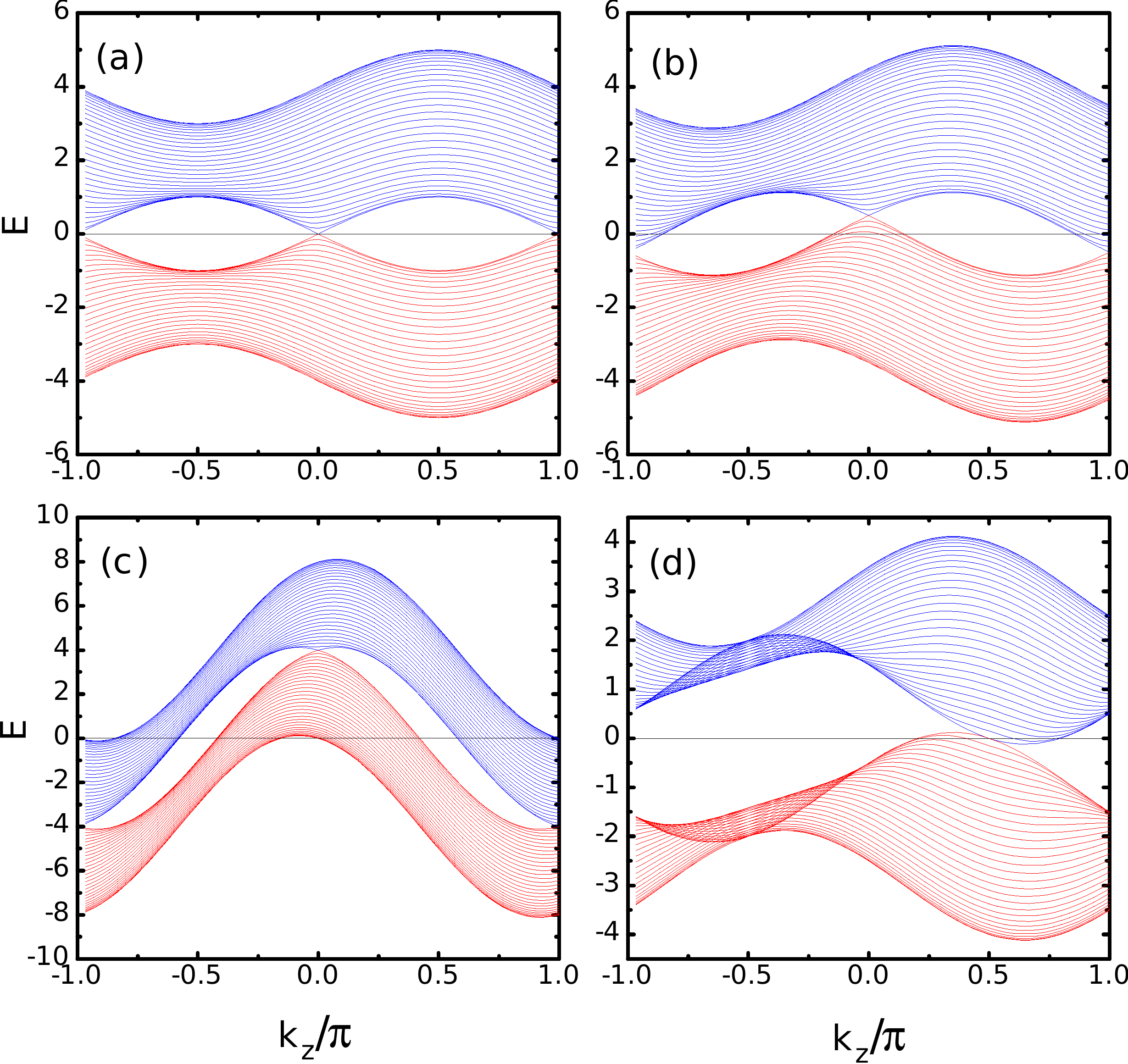}
\caption{Band energies as functions of $k_z$, with $k_x=k_y$, for $t_{so}=1$, $m=0$, and $t_1=0$ (a), 0.5 (b), 4 (c). In (d), 2 nodes merge at $k_z=\pi/2$ when $t_{so}=1$, $m=-1$, and $t_1=0.5$.}
\label{fig:band}
\end{figure}

In addition to the \emph{momentum} separation, one can also alter the \emph{energy} separation between the two nodes with the parameter $t_1$. For example, assuming $m=0$, then there are two nodes at $\mathbf{k}=(0,0,0)$ and $(0,0,-\pi)$. If $t_1=0$, then both nodes have zero energy. If $t_1\ne 0$, then the two nodes move to energies $t_1$ and $-t_1$, but their locations in momentum space remain unchanged. The band energies for some typical values of $t_1$ and $m$ are shown in Fig.~\ref{fig:band} for reference. We note that, the tuning of $t_1$ does not alter the positions of the nodes in momentum space only in this special case with $m=0$. For other values of $m$, the two separations in energy and in momentum are not entirely independent in the two-band model. In the following, we focus mainly on the case with $m=0$ for simplicity.

From now on, we set $t_{so}=1$ for convenience. The calculations below are done with large lattices (typically $800^3$ lattice sites) to eliminate finite-size effect. Within numerical accuracy, the calculated CME coefficient in the {\it static} limit [calculated using either Eq.~\eqref{chmag2} or Eq.~\eqref{chmag3}] is {\it always zero} for all ranges of the model parameters studied. Thus we only show the CME coefficient in the uniform limit [i.e., Eq.~\eqref{chmag1}] in the discussion below. The Hall conductivity $\sigma_H$ is evaluated by Eq.~\eqref{AHE}.

\subsection{Changing the energy separation between nodes}

\begin{figure}
\includegraphics[width=3.2in]{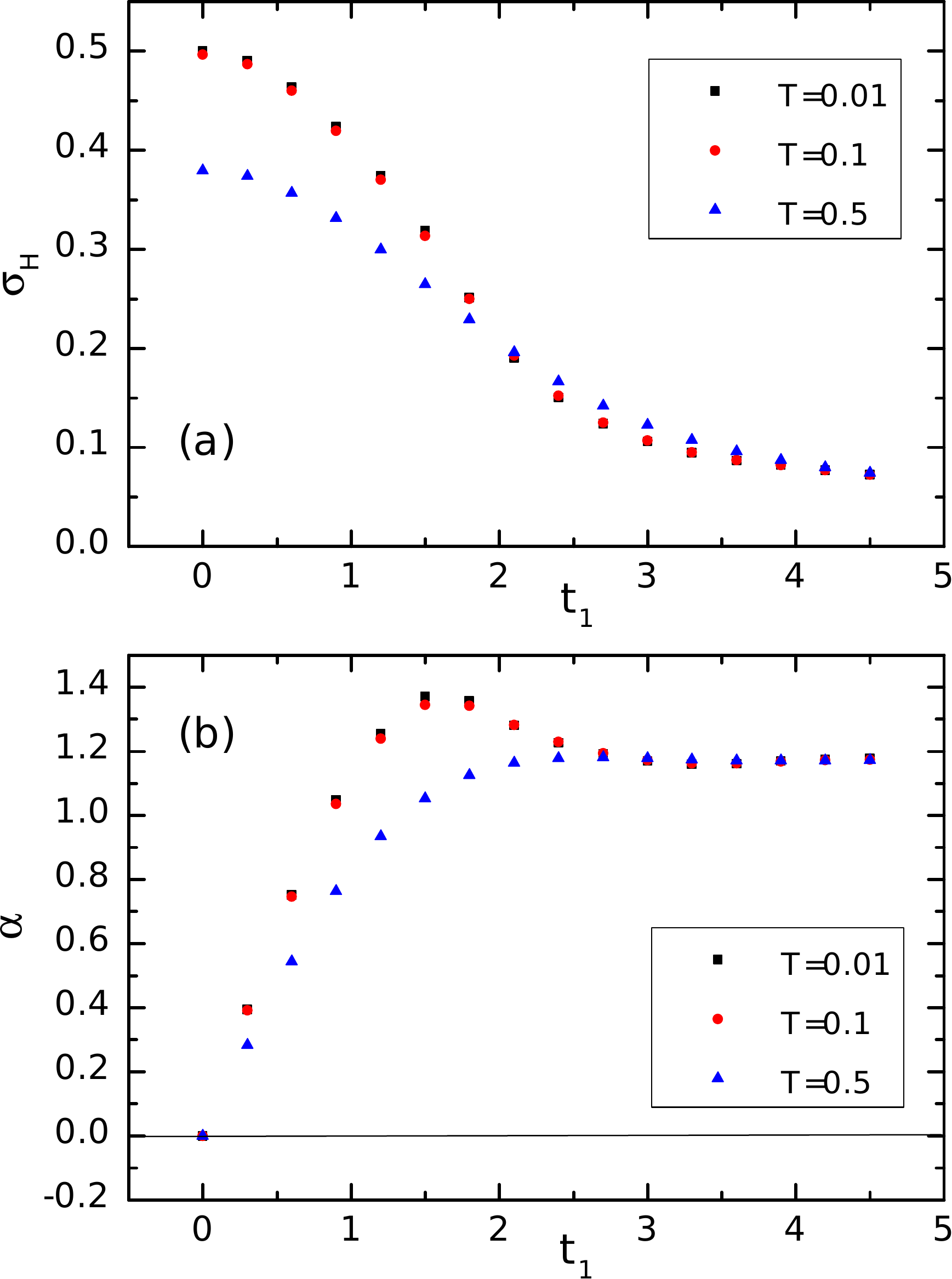}
\caption{(a) Hall conductivity $\sigma_H$ in units of $e^2/h$ and (b) CME coefficient $\alpha$ in units of $e^2/h^2$ in the uniform limit [i.e., Eq.~\eqref{chmag1}] as functions of $t_1$ at three different temperatures, $T=0.01$, 0.1, and 0.5. Here, $t_{so}=1$, $m=0$, and $\mu=0$. } \label{fig:t1_dep}
\end{figure}

In Fig.~\ref{fig:t1_dep}, we show the Hall conductivity $\sigma_H$ and the CME coefficient $\alpha$ as functions of $t_1$ for $m=0$ at a fixed chemical potential $\mu=0$. Recall that for $m=0$, the two Weyl nodes are located at $\mathbf{k}=(0,0,0)$ and $(0,0,-\pi)$ and the energy separation between them is $\triangle\epsilon=2t_1$. When $t_1=0$ and $\mu=0$, all states in the lower energy band are occupied at zero temperature [see Fig.~\ref{fig:band}(a)]. Since the Dirac string connecting the two Weyl nodes now span half of the $k_z$-axis, and each 2D cross section of the BZ that intersects the Dirac string would give a quantized Hall conductivity $e^2/h$, we have $\sigma_H=e^2/2h$. In Fig.~\ref{fig:band}, one observes that the two nodes separate in energy with increasing $t_1$. Due to the change in the electron populations around these two nodes, there is a gradual decrease in $\sigma_H$ as shown in Fig.~\ref{fig:t1_dep}(a). The curve goes asymptotically as $\sigma_H(t_1)\sim (1/\pi t_1)(e^2/h)$ at low temperatures. Besides, we find little difference between the results calculated from temperatures $T=0.01$ and $T=0.001$ (not shown), so further decrease in the temperature results in little change.

On the other hand, when $t_1=0$, there is no CME, since the two nodes have the same energy. When $t_1$ increases, as shown in Fig.~\ref{fig:t1_dep}(b), the CME coefficient $\alpha$ becomes nonzero. For small $t_1$ and at low temperatures, $\alpha(t_1)$ is roughly proportional to the energy separation $\triangle\epsilon=2t_1$ between the two Weyl nodes. This is consistent with the linear dependence predicted based on linearized effective models.~\cite{Zyuzin_Burkov2012,Goswami_Tewari2013,
Stephanov_Yin2012,Son_Yamamoto2012,Zyuzin_etal2012} However, when $t_1\gtrsim 1$ such that both energies of the Weyl nodes lie far away from the chemical potential $\mu=0$, the dependence of the CME coefficient on $t_1$ becomes nonlinear. Interestingly, the value of $\alpha$ saturates eventually for much larger $t_1$. Upon closer examination, we find that this is roughly due to the balance between the velocity $v_z\propto t_1$ and the Berry curvature $\Omega^z\propto 1/t_1$ at large $t_1$. (The latter fact also gives $\sigma_H\propto 1/t_1$ shown earlier.) Fig.~\ref{fig:t1_dep}(b) is one of the main results of this paper. It shows that the chiral magnetic effect does exist if the correct dc limit is taken. Besides, it also provides a guide to the general behaviour of the CME coefficient upon adjusting the energy separation of the two Weyl nodes.

We note that it is hard to understand the saturated value of the CME coefficient from the perspective of the chiral anomaly. As shown in Fig.~\ref{fig:band}(c), the chemical potential $\mu=0$ lies far away from the energies of the two Weyl nodes for large $t_1$. In this case, the linearized effective models around the Weyl nodes are not appropriate and the concept of chirality becomes ambiguous there. Therefore, it is the Berry curvature, rather than the chiral anomaly, that provides a better understanding of this effect.

\begin{figure}
\includegraphics[width=3.2in]{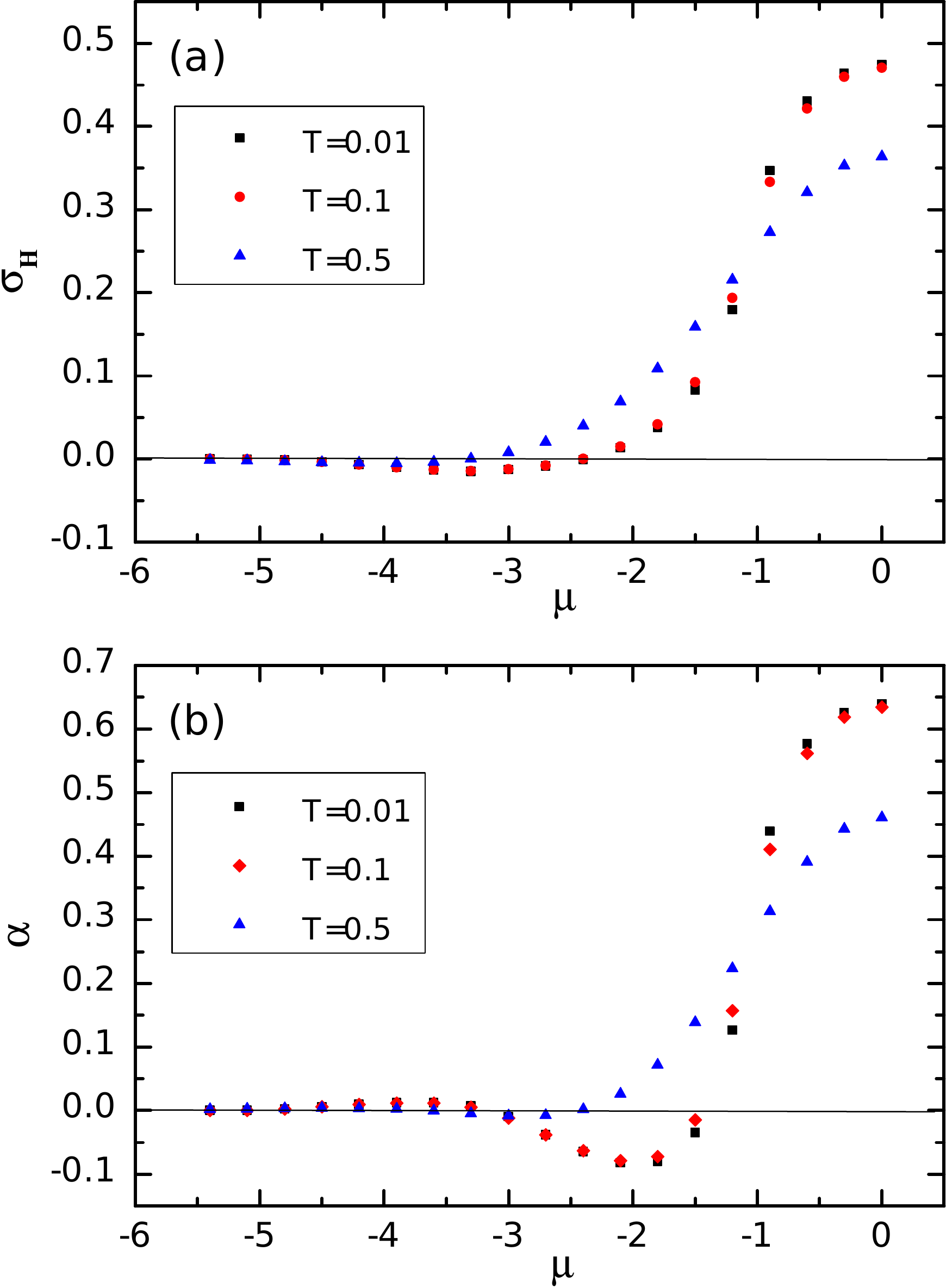}
\caption{(a) Hall conductivity $\sigma_H$ in units of $e^2/h$ and (b) CME coefficient $\alpha$ in units of $e^2/h^2$ in the uniform limit [i.e., Eq.~\eqref{chmag1}] as functions of the chemical potential $\mu$ at three different temperatures, $T=0.01$, 0.1, and 0.5. Here, $m=0$, $t_{so}=1$, and $t_1=0.5$.}
\label{fig:mu_dep}
\end{figure}

\subsection{Changing the chemical potential}

In Fig.~\ref{fig:mu_dep}, we show the Hall conductivity $\sigma_H$ and the CME coefficient $\alpha$ as functions of the chemical potential $\mu$ for $m=0$ and $t_1=0.5$. The two Weyl nodes are located at $\mathbf{k}=(0,0,0)$ and $(0,0,-\pi)$, and their energies are $t_1$ and $-t_1$. When $\mu=0$, the Hall conductivity $\sigma_H$ is slightly less than $e^2/2h$. From Fig.~\ref{fig:band}(b), we find that, as one decreases the chemical potential, the upper band will eventually become empty and electrons in the lower band will be populated less and less. Thus $\sigma_H$ becomes smaller as $\mu$ decreases from 0, as shown in Fig.~\ref{fig:mu_dep}(a). For the present case of $m=0$, the minimum of the lower band is located at $\mathbf{k}_\mathrm{min}=(\pi,\pi,\pi-\tan^{-1}(1/t_1))$ with the minimum energy $E_\mathrm{min}=-\sqrt{1+(t_1)^2}-4$, which gives $E_\mathrm{min}\simeq-5.1$ for $t_1=0.5$. Therefore,  $\sigma_H=0$ if $\mu\leq-5.1$, when the lower band also becomes empty. Notice that the Hall conductivity can be slightly negative when $\mu<-2.5$ at low temperatures. This is due to the distribution of negative Berry curvature at lower band energies.

The CME coefficient in the uniform limit as a function of the chemical potential $\mu$ is presented in Fig.~\ref{fig:mu_dep}(b). It behaves quite similarly to that of $\sigma_H$, since their dependence on $\mu$ is mainly controlled by the population of electrons. In general, the decrease in $\alpha$ is not monotonic, since the electron population and the Berry curvature within the BZ do not vary monotonically as the chemical potential changes. It eventually approaches zero when the lower band is completely empty (i.e., $\mu\leq-5.1$).

We have also calculated $\sigma_H$ and $\alpha$ for negative values of $t_1$ and positive values of $\mu$. It is found that $\sigma_H(-t_1)=\sigma_H(t_1)$ and $\alpha(-t_1)=-\alpha(t_1)$. Also, $\sigma_H(-\mu)=\sigma_H(\mu)$ and $\alpha(-\mu)=\alpha(\mu)$. These are not the most general features of $\sigma_H$ and $\alpha$, but are true for the two-band model studied.

\subsection{Changing the magnetization}

In this subsection, the dependence of the Hall conductivity $\sigma_H$ and the CME coefficient $\alpha$ on the magnetization $m$ is discussed. Here we take the chemical potential $\mu=0$.

We begin with the case of $t_1=0$ and temperature $T=0$. In this simpler case, the Weyl nodes (if they exist) all have zero energy and there is thus no CME. That is, $\alpha$ will be zero for the whole range of $m$. On the other hand, the Hall conductivity $\sigma_H$ depends nontrivially on $m$. For $t_1=0$ and $\mu=0$, all states in the lower energy band are occupied at zero temperature, and the Hall conductivity $\sigma_H$ is thus proportional to the length of the Dirac string $\Delta k_z$. As being explained at the beginning of this section, the value of $m$ determines the distance between a pair of Weyl nodes (see Fig.~\ref{fig:node}), and the dependence on $m$ of the Hall conductivity $\sigma_H$ can be easily understood.

When $m>1$ ($t_{so}=1$), two bands are separated by an energy gap and there is no Weyl node. Therefore, $\sigma_H=0$. If $m$ is decreased from 1 to $-1$, the length of the Dirac string $\Delta k_z$ increases from zero to $2\pi$, as shown in Fig.~\ref{fig:node}(a). Therefore, $\sigma_H$ will monotonically increase from zero to its maximum value of $e^2/h$. When one reduces $m$ from $-1$ to $-3$, two anti-Dirac strings along the $k_z$ axis appear, as seen from Fig.~\ref{fig:node}(b). So the Hall conductivity decreases as the nodes move away from each other, and it reaches the minimum value of $-e^2/h$ when $m=-3$. When $m$ decreases from $-3$ to $-5$, a Dirac string appears in Fig.~\ref{fig:node}(c), which leads to an increase of the Hall conductivity. Finally, the negative $\sigma_H$ is compensated to reach the value of zero when $m=-5$. Beyond that, the system is a trivial insulator with $\sigma_H=0$.

\begin{figure}
\includegraphics[width=3.2in]{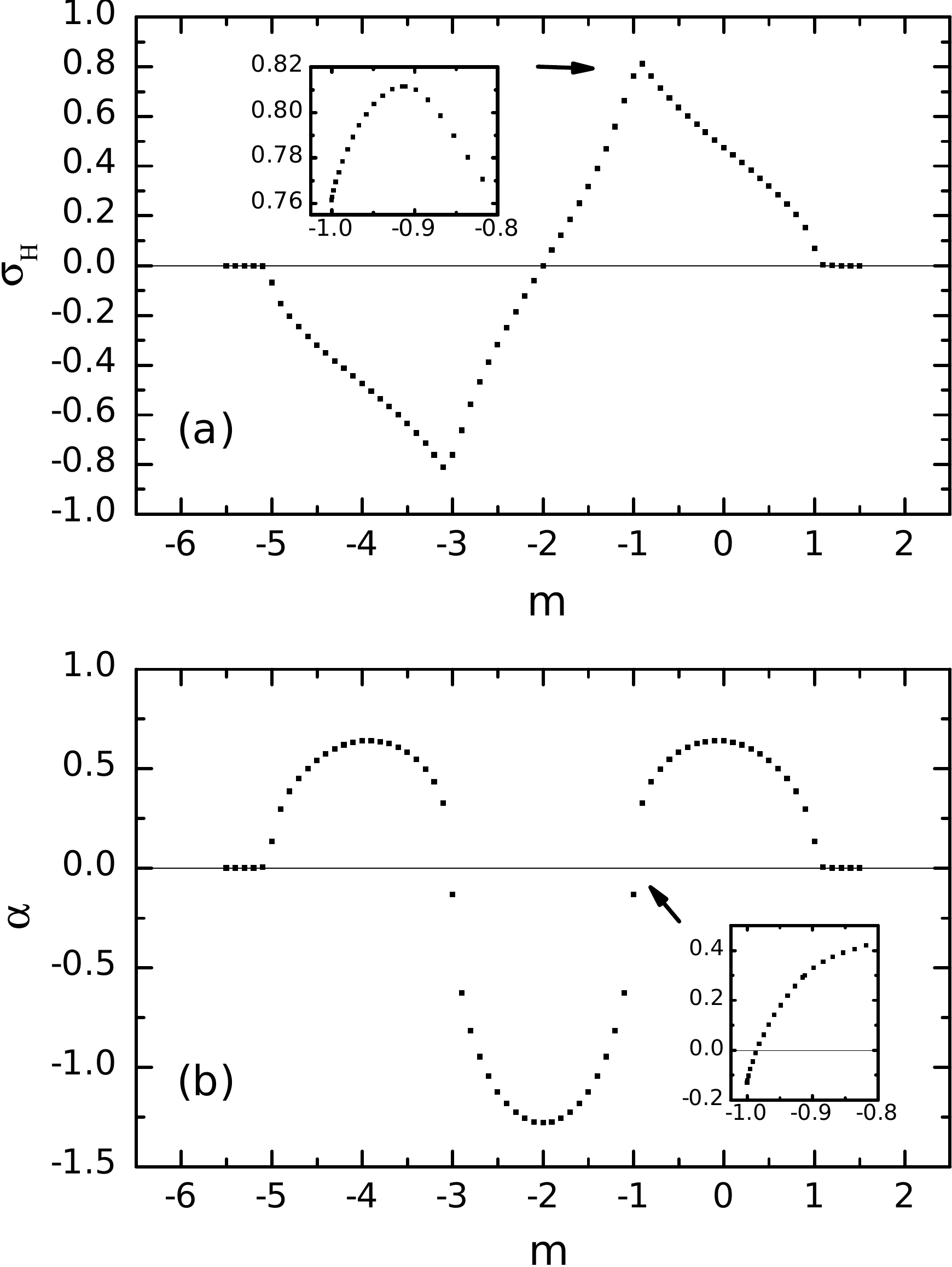}
\caption{(a) Hall conductivity $\sigma_H$ in units of $e^2/h$ as a function of the magnetization $m$. Inset: details of the peak near $m=-1$. (b) CME coefficient $\alpha$ in units of $e^2/h^2$ in the uniform limit [i.e., Eq.~\eqref{chmag1}] as a function of $m$. Inset: details of the crossing of zero near $m=-1$. Here, $t_{so}=1$, $t_1=0.5$, $\mu=0$, and $T=0.01$. The system is a trivial insulator with zero CME coefficient when $m>t_{so}$ and $m<-5t_{so}$. It is in an anomalous Hall phase when $-5t_{so}<m<t_{so}$. Notice that the positions of the peak of $\sigma_H$ in (a) and the zero of $\alpha$ in (b) do not coincide. This difference grows for larger $t_1$.}
\label{fig:m_dep}
\end{figure}

We now come back to the case when $t_1\ne 0$. The dependence of the Hall conductivity $\sigma_H$ on the magnetization $m$ for $t_1=0.5$ at low temperature $T=0.01$ is presented in Fig.~\ref{fig:m_dep}(a). One can see that $\sigma_H$ is anti-symmetric with respect to $m=-2$. The general behaviour looks similar to that described above for the $t_1=0$ case. However, the Weyl nodes move away from zero energy for the present case of $t_1=0.5$, and one can not fill only the lower band even when $\mu=0$. Therefore, the peak values of $\sigma_H$ is no longer quantized at $\pm e^2/h$. Besides, one now has smooth peaks located away from $m=-1$ and $-3$ [see the inset of Fig.~\ref{fig:m_dep}(a) around $m=-1$]. Such a deviation becomes quite apparent if one chooses a larger value of $t_1$.

On the other hand, the dependence of the CME coefficient $\alpha$ on  $m$ for $t_1=0.5$ and $T=0.01$ is plotted in Fig.~\ref{fig:m_dep}(b), which is symmetric with respect to $m=-2$. We note that $\alpha$ changes sign near, but not at, $m=-1$ and $m=-3$ [see the inset of Fig.~\ref{fig:m_dep}(b) around $m=-1$], even though the Weyl nodes merge in the 3D BZ for these special values of $m$, as shown in Fig.~\ref{fig:node}. This is due to the fact that, when $t_1\ne 0$, the two bands overlap near the chemical potential $\mu=0$, before the merge of the Weyl nodes at $k_z=\pi/2$ [see Fig.~\ref{fig:band}(d)]. The deviation of the locations of $\alpha=0$ from these special values of $m$ for node merging becomes quite apparent at a larger value of $t_1$. Since the parameter $t_1$ has nothing to do with the topology of the ground state, such a sign change of $\alpha$ is not related to a quantum phase transition, but is only a result of the distortion of the electronic structure.

Also, notice that there is no chiral magnetic effect when the lower band is completely filled ($m>1$ or $m<-5$). That is, a filled band cannot carry an electric current, which is consistent with conventional wisdom.

\section{Discussion and Conclusion}

From this investigation, the following lessons have been learned. First,
the choice of the ultra-violet cutoff can be crucial in calculating quantities for unbounded, linearized effective theories. In the present case, energy cutoff should be used and then the correct result of null equilibrium current will be reached.~\cite{Zhou_etal2013,Basar_etal2014,Landsteiner2014}
Therefore, one should be cautious in deriving results by using unbounded, linearized effective theories.

Second, different limiting procedures usually correspond to different experimental arrangements and thus different physical results.
The correct dc transport quantities, say, the CME coefficient studied here, should be obtained under the uniform limit.~\cite{Chen_etal2013} Confusing conclusions can be drawn under the wrong type of limit.

Besides, we note that the non-analyticity in $\alpha(\mathbf{q},\omega)$ poses a serious issue regarding the validity of the \emph{local} Chern-Simons term in the effective action of Weyl semimetals. Usually, the fascinating electromagnetic properties of Weyl semimetals is described by a radiatively induced Chern-Simons term in the effective action after integrating out fermions,~\cite{Hosur_Qi2013}
\begin{equation}\label{S_eff}
S_\textrm{eff}=-\frac{e^2}{4\pi^2 \hbar}\int d^3x dt\; b_\sigma \epsilon^{\mu\nu\rho\sigma} A_\mu \partial_\rho A_\nu \; .
\end{equation}
The coefficients $b_\sigma$ can be obtained from the derivatives of an ``axion" field $\theta(\mathbf{x},t)$. In the frequency-momentum domain, the effective action reads as
\begin{equation}
S_\textrm{eff} = \int \frac{d^4q}{(2\pi)^4}\; A_{\mu}(-q) \Pi_{\mu\nu}(q) A_{\nu}(q)
\end{equation}
with
\begin{equation}
\Pi_{\mu\nu}(q)= i\epsilon^{\mu\nu\rho\sigma} q_\rho \alpha_{\sigma}(q) \; ,
\end{equation}
which is nothing but the generalization of Eq.~\eqref{Pi_anti_def}. That is, the coefficients $b_\sigma$ are proportional to $\alpha_{\sigma}(q)$ in the limit of zero frequency and momentum. However, as shown above, $\alpha_{\sigma}(q)$ are not analytic at $(\mathbf{q}, \omega)=(\mathbf{0}, 0)$. This indicates that we are not always allowed to expand the effective action as a series of local terms. That is, we do not have a \emph{bona fide} induced local Chem-Simons term.

Finally, a cautionary remark is given. In this work, clean and infinite systems are presumed. However, real samples are always finite in sizes (with possible surface states), accompanied by disorders, and connected to external reservoirs. It is known that the Kubo-formula results for usual electric transport remain mostly reliable in realistic situations. Nevertheless, in the present case, whether these complications would qualitatively alter the conclusion in this paper (and nullify the CME) remains an open question. It is an important issue worthy of further investigations.

In conclusion, we address the issue of the existence of the CME carefully. For Weyl semimetals described by a generic two-band model, different expressions of the CME coefficient are obtained under different limiting procedures. It shows that the CME coefficient behaves non-analytically in the zero frequency-momentum limit. Employing an explicit form of the two-band model, we show that the chiral magnetic current in an equilibrium state under the static limit vanishes as expected. Nevertheless, the CME does exist under the correct dc limit (i.e., the uniform limit) and its proportional coefficient exhibits nontrivial dependence on various system parameters. Furthermore, even when the linearization of energy dispersion is not valid, such that chirality may not be well defined, the CME can still be finite. It shows that this effect does not crucially depend on the chiral anomaly. Our work clarifies some subtleties in calculating the CME coefficient. It should be of help to future researchers studying related effects of Weyl semimetals and other nodal materials.

\begin{acknowledgments}
The authors would like to thank A. A. Burkov, N. Yamamoto, and J.-Y. Chen for discussions.
M.C.C. and M.F.Y. acknowledge the support from the National Science
Council of Taiwan under Grant Nos. NSC 102-2112-M-003-005-MY3 and NSC 102-2112-M-029-004-MY3, respectively.
\end{acknowledgments}

\appendix
\section{momentum cutoff versus energy cutoff}
\label{cutoff}

Here the linearized effective model of Weyl semimetal introduced in Ref.~\onlinecite{Zyuzin_Burkov2012} is employed to illustrate the crucial role played by the ultra-violet cutoff. We show that different regularization schemes can lead to distinctive results of the CME.

In the presence of a static uniform magnetic field in the $z$ direction,
$\mathbf{B} = B\hat{\mathbf{z}}$, using Landau-level ladder operators $a$ and $a^\dag$, the Hamiltonian becomes [see Eq.~(64) in  Ref.~\onlinecite{Zyuzin_Burkov2012}]
\begin{equation}
H = \frac{\omega_B}{\sqrt{2}} \tau^z (\sigma^+ a + \sigma^-
a^\dag) + \tau^z \sigma^z k_z + \tau^z b_0 - \tau^x\Delta \; ,
\end{equation}
where $\omega_B=\sqrt{eB}$ is the inverse of the magnetic length and $\Delta$ is a uniform time-independent node-hybridizing potential. The Pauli matrices $\boldsymbol{\sigma}$ and $\boldsymbol{\tau}$ act on the real spin and the surface-pseudospin degrees of freedom, respectively. The energy separation between the two Weyl nodes is $2b_0$, and $k_z$ is the momentum in the $z$ direction.

Diagonalizing this Hamiltonian gives the following Landau-level dispersions
\begin{equation} \label{eq:En}
\epsilon_{n s \alpha} = s \sqrt{\left(\sqrt{2 \omega_B^2 n +
k_z^2} + \alpha b_0\right)^2 + \Delta^2}
\end{equation}
for $n\geq 1$, while the $n=0$ Landau-level dispersions are
\begin{equation} \label{eq:E0}
\epsilon_{0 \alpha} = \alpha \sqrt{(k_z - b_0)^2 + \Delta^2},
\end{equation}
where $s,\alpha=\pm$. We note that $\epsilon_{0 \alpha}$ have no inversion symmetry with respect to $k_z=0$ when $b_0\neq 0$.

The equilibrium current in response to the applied magnetic field can be calculated by the following formula~\cite{Fukushima_etal2008}
\begin{equation} \label{eq:jz}
J^z = - \frac{e^2 B}{2\pi \hbar^2} \int_{-\Lambda}^{\Lambda}
\frac{d k_z}{2 \pi} \frac{d}{d k_z}\left( \epsilon_{0 -} + \sum_{n=1}^{\infty} \sum_{\alpha=\pm} \epsilon_{n - \alpha} \right),
\end{equation}
where $d\epsilon_{n}/dk_z$ is the $z$-component electron velocity in the $n$-th Landau level. To subtract the contribution from the infinite Dirac sea, certain ultra-violet cutoff must be introduced, which will be taken to infinity at the end of the calculation. Usually the momentum cutoff $\Lambda$ symmetric about $k_z=0$ is considered, as shown in Eq.~\eqref{eq:jz}. However, another choice of cutoff can produce a different result, as discussed below.

Since $\epsilon_{n s \alpha}$ for $n \geq 1$ are even functions of
$k_z$, only the $n=0$ Landau level can give nonzero contribution to
$J^z$. Thus we obtain
\begin{eqnarray}
J^z&=&- \frac{e^2 B}{2\pi \hbar^2} \int_{-\Lambda}^{\Lambda} \frac{d k_z}{2 \pi} \frac{d \epsilon_{0 -}}{d k_z}
\nonumber \\
&=&\frac{e^2 B}{4\pi^2 \hbar^2} \left[(\Lambda - b_0) - (\Lambda + b_0) \right] = - \frac{b_0 e^2}{2\pi^2 \hbar^2} B \; ,
\end{eqnarray}
where the second line is true in the limit $\Lambda/ \Delta
\rightarrow \infty$. This gives the CME coefficient $\alpha=b_{0}e^{2}/2\pi^2 \hbar^2$, which is the value reported in some literature.~\cite{Stephanov_Yin2012,
Son_Yamamoto2012,Zyuzin_Burkov2012,Grushin2012,Goswami_Tewari2013}

However, as emphasized in Refs.~\onlinecite{Basar_etal2014} and \onlinecite{Landsteiner2014}, because of the finite band width, there always exists band minima in a solid, and one should use \emph{energy} cutoff instead. If we now subtract the contributions from the states with energies lower than the energy cutoff $-\epsilon_0$ (where $\epsilon_0 \gg
b_0$, $\Delta$), according to the Landau-level dispersions in
Eqs.~\eqref{eq:En} and~\eqref{eq:E0}, the corresponding region of the $k_z$-integration becomes
$$-\Lambda_n  \leq  k_z  \leq  \Lambda_n$$
with $\epsilon_{n - \alpha}(\pm\Lambda_n)=-\epsilon_0$ for $n \geq
1$; while for $n=0$,
$$\Lambda_-   \leq  k_z  \leq  \Lambda_+ $$
with $\Lambda_\pm \equiv \pm\sqrt{\epsilon_0^2 - \Delta^2} + b_0$.
Note that $\Lambda_\pm$ are not symmetric with respect to $k_z=0$ when $b_0\neq 0$. Again, only the $n=0$ Landau level contributes to $J^z$, and it gives
\begin{eqnarray}
J^z &=& -\frac{e^2 B}{2\pi \hbar^2}
\int_{\Lambda_-}^{\Lambda_+} \frac{d k_z}{2 \pi} \frac{d
\epsilon_{0 -}}{d k_z} \nonumber \\
&=& -\frac{e^2 B}{4\pi^2 \hbar^2}
\left[ (-\epsilon_0) - (-\epsilon_0) \right] = 0 \; .
\end{eqnarray}
That is, there is no chiral magnetic effect. This conclusion is consistent with the viewpoint that there is no equilibrium current in the static limit.~\cite{Chen_etal2013,Mahan,Thouless,Luttinger1964}

In short, the choice of cutoff is important in the evaluation of the electric current for the unbounded effective models. If we use \emph{symmetric} momentum cutoff (say, $k_z=\pm\Lambda$) for the $n=0$ Landau level, as employed in Ref.~\onlinecite{Zyuzin_Burkov2012}, a nonzero current will be found. This may be the reason why nonzero currents are observed in the case of  a lattice model when artificial \emph{symmetric} momentum cutoffs are introduced.~\cite{Vazifeh_Franz2013} However, as explained in Refs.~\onlinecite{Basar_etal2014} and \onlinecite{Landsteiner2014}, the method of  choice in condensed matter should be the energy cutoff, which corresponds to \emph{asymmetric} momentum cutoff (that is, $k_z=\Lambda_\pm$ in the present case), and then the expected zero current in the static limit can be recovered.

\section{semiclassical wave-packet dynamics in the presence of a time-dependent magnetic field}
\label{semi}

In this appendix, the correction term in the distribution function related to a time-dependent magnetic field is derived. This derivation is based on the discussions in Ref.~\onlinecite{Son_Yamamoto2013}. However, here we go beyond the scope of (effective) models with linear dispersion.

Under weak electric and magnetic fields, the semiclassical equations of motion for
a Bloch electron are~\cite{Xiao_etal2010}
\begin{subequations}
\label{EOM}
\begin{align}
\dot{\mathbf{x}} &= \frac{1}{\hbar}
\frac{\partial\tilde{\varepsilon}_n}{\partial\mathbf{k}}
- \dot{\mathbf{k}} \times \mathbf{\Omega}_n(\mathbf{k}) \; , \\
\dot{\mathbf{k}} &= -\frac{e}{\hbar} \tilde{\mathbf{E}} -\frac{e}{\hbar}\dot{\mathbf{x}}
\times \mathbf{B} \; .
\end{align}
\end{subequations}
Here $\mathbf{\Omega}_n(\mathbf{k})$ is the Berry curvature of Bloch states in the $n$-th band, $\tilde{\mathbf{E}}\equiv\mathbf{E}+(1/e)\partial\tilde{\varepsilon}_n/\partial\mathbf{x}$, and $\tilde{\varepsilon}_n=\varepsilon_n(\mathbf{k})-\mathbf{m}_n(\mathbf{k})\cdot\mathbf{B}$ is the band energy $\varepsilon_n(\mathbf{k})$ with a correction from the orbital magnetic moment $\mathbf{m}_n(\mathbf{k})$.\cite{chang1996,sundaram1999}

Solving Eqs.~(\ref{EOM}), we get
\begin{align}\label{sol_EOM}
\dot{\mathbf{x}} &= \frac{1}{D_n(\mathbf{k})}
  \left[ \tilde{\mathbf{v}}_n + \frac{e}{\hbar}\tilde{\mathbf{E}}\times\mathbf{\Omega}_n(\mathbf{k})
  +\frac{e}{\hbar}(\mathbf{\Omega}_n(\mathbf{k})\cdot\tilde{\mathbf{v}}_n)\mathbf{B}
  \right] \; ,\nonumber \\
\dot{\mathbf{k}} &= \frac{1}{D_n(\mathbf{k})}
  \left[ -\frac{e}{\hbar}\tilde{\mathbf{E}} -\frac{e}{\hbar}\tilde{\mathbf{v}}_n\times\mathbf{B}
  -\frac{e^2}{\hbar^2}(\tilde{\mathbf{E}}\cdot\mathbf{B})\mathbf{\Omega}_n(\mathbf{k}) \right] \; ,
\end{align}
where $\tilde{\mathbf{v}}_n=
(1/\hbar)\partial\tilde{\varepsilon}_n/\partial\mathbf{k}$ is the group velocity of the Bloch electrons and $D_n(\mathbf{k})\equiv 1+(e/\hbar)\mathbf{B}\cdot\mathbf{\Omega}_n(\mathbf{k})$ is the Berry curvature correction to the density of states in phase space.~\cite{Xiao_etal2005}

The kinetic equation governing the time evolution of the electron distribution function $n_\mathbf{k}$ is
\begin{equation}\label{kineq}
  \frac{\partial n_\mathbf{k}}{\partial t}
  + \dot{\mathbf{x}}\cdot\frac{\partial n_\mathbf{k}}{\partial\mathbf{x}}
    + \dot{\mathbf{k}}\cdot\frac{\partial n_\mathbf{k}}{\partial \mathbf{k}}
  = 0 \; .
\end{equation}
Substituting Eq.~\eqref{sol_EOM} to Eq.~\eqref{kineq} and keeping terms up to first order in $\mathbf{E}$ and $\mathbf{B}$, one has
\begin{equation}
\frac{\partial n_\mathbf{k}}{\partial t}
+\mathbf{v}_n \cdot \frac{\partial n_\mathbf{k}}{\partial\mathbf{x}}
+\frac{1}{\hbar}
\left[ -e\mathbf{E} +
\frac{\partial(\mathbf{m}_n(\mathbf{k})\cdot\mathbf{B})}{\partial\mathbf{x}} \right]
\cdot \frac{\partial n_\mathbf{k}}{\partial\mathbf{k}} = 0 \; ,
\end{equation}
where $\mathbf{v}_n=(1/\hbar)\partial\varepsilon_n/\partial\mathbf{k}$ is the group velocity of a Bloch electron in \emph{zero} field.
We note that $\partial n_\mathbf{k}/\partial\mathbf{x}$ and
$\partial\tilde{\varepsilon}_\mathbf{k}/\partial\mathbf{x}$ are at least of linear order in $\mathbf{E}$ and $\mathbf{B}$. Moreover, the $\tilde{\mathbf{v}}_n\times\mathbf{B}$ term gives no contribution
since $(\tilde{\mathbf{v}}_n\times\mathbf{B})\cdot\tilde{\mathbf{v}}_n=0$.

The distribution function $n_\mathbf{k}$ can be decomposed as
\begin{eqnarray}
n_\mathbf{k} &=& \tilde{f}_n(\mathbf{k}) + \delta n_\mathbf{k}
\nonumber \\
&\simeq& f_n(\mathbf{k})
+ \left( -\mathbf{m}_n(\mathbf{k})\cdot\mathbf{B}
\;\frac{\partial f_n}{\partial\varepsilon_n}
+ \delta n_\mathbf{k} \right) \; ,
\end{eqnarray}
where $\tilde{f}_n(\mathbf{k})$ and $f_n(\mathbf{k})$ are the Fermi-Dirac distributions for the band energies $\tilde{\varepsilon}_n=\varepsilon_n(\mathbf{k})-\mathbf{m}_n(\mathbf{k})\cdot\mathbf{B}$ and $\varepsilon_n(\mathbf{k})$, respectively. The additional field-induced correction $\delta n_\mathbf{k}$ in the distribution function thus satisfies the differential equation,
\begin{equation}
\left(\frac{\partial}{\partial t} + \mathbf{v}_n \cdot \frac{\partial}{\partial\mathbf{x}} \right)
\delta n_\mathbf{k}
= \left[ \frac{\partial(\mathbf{m}_n(\mathbf{k})\cdot\mathbf{B})}{\partial t}
+ e\mathbf{E}\cdot\mathbf{v}_n \right]
\frac{\partial f_n}{\partial\varepsilon_n} \; .
\end{equation}

Suppose $\mathbf{E}$, $\mathbf{B}\propto \exp{\{i(\mathbf{q}\cdot\mathbf{x}-\omega t)\}}$ and $\delta n_\mathbf{k}$ varies in space and time in the same manner, then the solution of $\delta n_\mathbf{k}$ is
\begin{equation}\label{nk}
\delta n_\mathbf{k} = \frac{1}{\omega-\mathbf{q}\cdot\mathbf{v}_n}
[ \omega\,\mathbf{m}_n(\mathbf{k})\cdot\mathbf{B} + ie\mathbf{E}\cdot\mathbf{v}_n ]
\,\frac{\partial f_n}{\partial\varepsilon_n}
\; .
\end{equation}
This result is basically identical to the combination of Eqs.~(95) and (96) in Ref.~\onlinecite{Son_Yamamoto2013} after transforming into the frequency-momentum domain. However, our expression is applicable even beyond the scope of (effective) models with linear dispersion.

Now it is clear that $\delta n_\mathbf{k}$ (and thus the distribution function $n_\mathbf{k}$) behaves differently under different orders of the zero frequency-momentum limits. For a static magnetic field (i.e., $\omega=0$) and without an electric field, we have $\delta n_\mathbf{k}=0$, and the distribution function in this static limit becomes
\begin{equation}\label{nk_static}
n_\mathbf{k}^\textrm{static} \simeq f_n(\mathbf{k})
-\mathbf{m}_n(\mathbf{k})\cdot\mathbf{B}
\;\frac{\partial f_n}{\partial\varepsilon_n}
\; .
\end{equation}
Employing this form of the distribution function, one can reproduce the expression of the CME coefficient obtained in previous literature.~\cite{Zhou_etal2013,Basar_etal2014,Landsteiner2014} That is, their result corresponds to that in the static limit.

On the other hand, a time-dependent but spatially uniform magnetic field (i.e., $\mathbf{q}=0$) will give a nonzero magnetic-field-induced correction $\delta n_\mathbf{k} =\mathbf{m}_n(\mathbf{k})\cdot\mathbf{B}
\;\partial f_n/\partial\varepsilon_n$. Therefore, the distribution function in the uniform limit becomes
\begin{equation}\label{nk_uniform}
n_\mathbf{k}^\textrm{uniform} \simeq f_n(\mathbf{k})  \; .
\end{equation}
Only after taking the additional correction term $\delta n_\mathbf{k}$ into account can one have the correct expression of the CME coefficient in the uniform limit.

In summary, we have shown that an extra contribution $\delta n_\mathbf{k}$ in the distribution function induced by a time-dependent magnetic field can lead to a result different from that in the static case. This gives the non-analyticity in the current response function found in the linear-response theory, as shown in Ref.~\onlinecite{Son_Yamamoto2013} for the Weyl model with linear energy dispersion.



\begin{thebibliography}{99}
\bibitem{note3}
Theoretically, to simulate the physics of gapless systems with energy levels arbitrarily close in energy, the thermodynamic limit (or the infinite-size limit) should be taken before $\omega\to 0$. Otherwise, the discrete nature of the energy levels due to the finite-size effect would forbid the transitions among levels in the zero-frequency limit and suppress the electron transport. Since a finite length scale
$L$ sets a minimum wave vector, say, $q_x\sim 1/L$, the limit of $L\to \infty$ before $\omega\to 0$ is transcribed to the uniform limit.
\end{thebibliography}
\end{document}